\documentstyle[12pt]{article}
\def\Lc#1{{\rm Lc}_{#1}}

\def\Li2{{\rm Li_2}}
\def\A{{\cal A}}
\def\L{{\cal L}}
\def\Lhat{\hat {\cal L}}
\def\T{{\cal T}}
\def\M{{\cal M}}
\def\P{{\cal P}}
\def\e{\epsilon}
\begin{document}
         %%%%%

\begin{titlepage}
\begin{flushright}
DTP/96/66\\
hep-ph/9609474\\
\end{flushright}
\vspace{1cm}
\begin{center}
{\Large\bf The One-loop QCD Corrections for $ \gamma^* \to Q\bar Q q\bar q$}\\
\vspace{1cm}
{\large
E.~W.~N.~ Glover and D.~J.~Miller\footnote{Address after 1 January 1997,
Rutherford Appleton Laboratory,  Chilton, Didcot, Oxon,
OX11 0QX, England}}\\
\vspace{0.5cm}
{\it
Physics Department, University of Durham,\\
Durham DH1~3LE,
England} \\
\vspace{0.5cm}
{\large September 1996}
\vspace{0.5cm}
\end{center}
\begin{abstract}
We present the first calculation of the
one-loop QCD corrections for the decay of an off-shell
vector boson with vector couplings into two pairs of quarks of equal
or unequal flavours keeping all orders in the number of colours.
These matrix elements are relevant for the calculation of the
next-to-leading order ${\cal O}(\alpha_s^3)$ corrections to
four jet production in electron-positron annihilation,
the production of a gauge boson accompanied by two jets in
hadron-hadron collisions and three jet production in deep inelastic
scattering. We compute the interference of one-loop
and tree level Feynman diagrams, and organise the matrix elements in terms of
combinations of scalar loop integrals that are well behaved in the limit of
vanishing Gram determinants.
The results are therefore numerically stable and ready to be implemented in
next-to-leading order
Monte Carlo calculations of the
$e^+e^- \to 4~{\rm jet}$,
$e^\pm p \to e^\pm + 3~{\rm jet}$ and
$p\bar p \to V + 2~{\rm jet}$  processes.

\end{abstract}
\end{titlepage}

An important ingredient in the confrontation of hadronic jet
data from high energy
particle collisions with theoretical calculations in perturbative QCD
is the calculation of parton level cross sections beyond leading order.
This is so for a variety of reasons.   First, the dependence on the
unphysical renormalisation and factorisation scales is usually reduced.
Second,
the addition of more particles into the final state makes the
calculation more sensitive
to the details of the experimental jet algorithm.  Third, the
final state phase space is enlarged so that lowest order constraints
on the jet transverse energy or rapidity are relaxed.  Together,
these improvements make for a more stable calculation that is
better matched to the experimental data.
Finally, the presence of large infrared logarithms can be detected and the
appropriate resummations performed.
In this paper, we present the first calculation
of the one-loop matrix elements for a
virtual gauge boson
decay into two pairs of equal or unequal flavour quarks
keeping all terms in the number of colours.
These matrix elements are one of the ingredients of the
next-to-leading order
calculation relevant for the processes
\begin{itemize}
\item $e^+e^- \to 4~{\rm jets}$
\item $e^\pm p \to e^\pm + 3~{\rm jets}$
\item $p\bar p \to V + 2~{\rm jets}$  with $V = W$ or $Z$,
\end{itemize}
which have been observed in abundance in experiments
at the major accelerators,
LEP at CERN, HERA at DESY and the TEVATRON at Fermilab.
In electron-positron collisions, applications include more precise
measurements of the colour casimirs $C_A/C_F$ and $T_F/C_F$ and
a reduction of the theoretical error in the extraction of the $WW$ cross
section near threshold.

In recent years there have been several significant technical advances in
one-loop Feynman diagram calculations,
some relying on insight from
supersymmetry and others inspired more directly by the superstring \cite{BDK3}.
Most important has been the continuation of the one-loop scalar pentagon
integral in 4-dimensions \cite{vNV}
to $D=4-2\epsilon$ dimensions \cite{BDK1,BDK2,EGY}.
At the same time, methods utilised in tree level matrix element calculations,
colour decompositions and helicity have been
taken over.
To make best use of the helicity method, relations have been provided
between
conventional dimensional regularisation,
where all polarisations and spinors are treated in $D$-dimensions,
and other schemes where parts of the calculations are performed in 4-dimensions
\cite{KST,CST}.
With these technological improvements, compact one-loop
helicity amplitudes for processes involving five partons
have been presented by Bern, Kosower and Dixon \cite{BDK5g,BDK2q3g} and
Kunszt, Signer and Trocsanyi \cite{KST4qg,Sig4qp,Signer},
while recently, one of the helicity amplitudes for the leading colour
part of the $Z \to q\bar q gg$ process has been evaluated \cite{BDKZ2q2g}.
Recently, Dixon and Signer have reported first numerical
results for the
leading colour contribution to the $e^+e^- \to 4~{\rm jet}$ rate
at next-to-leading order \cite{DS}.
As expected, the agreement between theory and experiment is significantly
improved.
Our results presented here can therefore provide a
numerical check of the one-loop
leading colour contribution from the four quark process.\footnote{
 Since this paper was
completed, Bern et al. \cite{BDKW} have presented analytic
results for all of the helicity amplitudes for the
$V \to Q\bar Q q\bar q$ process.
Their approach is completely different from ours and, using the
known relations between amplitudes calculated in dimensional
regularisation and the 4-D helicity scheme \cite{KST,CST},
a numerical comparison will provide a powerful check of our results.}

For our calculation involving a massive external gauge boson,
we choose to compute the interference between the tree-level
and one-loop amplitudes directly.
The resulting
`squared' matrix elements involve only dot products of the particle
four-momenta in addition to the normal logarithms and dilogarithms
encountered in one-loop
Feynman diagram calculations.
As a consequence, we can immediately
eliminate all integrals more complicated than
a scalar pentagon, although some box tensor-integrals with three loop momenta
remain.
We have kept all quantities in $D$-dimensions as in conventional
dimensional regularisation, but have
checked that the 't Hooft-Veltman scheme \cite{HV}
gives identical results \cite{KST,CST}.
Throughout we reorganise the tensor loop integrals
as combinations of scalar integrals using a momentum decomposition
\cite{PV} (rather than the reciprocal momentum basis of \cite{OV,EGY,Signer}).
The usual problem with this approach is the occurence and proliferation
of Gram determinants ($\Delta$) in the denominator which both increases
the size of the algebraic expressions and introduces numerical
instabilities as $\Delta \to 0$.
We have avoided this by introducing combinations of
scalar loop integrals in groups that are finite in the limit that the
Gram determinants vanish \cite{BDK5g,BDK2q3g,KST4qg,Signer,CGM}.
The matrix elements, written in terms of these finite functions,
are then obviously well behaved in the same limits.
These groupings are derived by differentiating the scalar integrals with
respect to the external kinematic factors or by considering them in
higher dimensions \cite{CGM}.
They combine the dilogarithms, logarithms and constants from different
scalar integrals in an extremely non-trivial way and, in doing so,
reduce the size of the expressions considerably.
However, these finite combinations are not linearly
independent (unlike the raw scalar integrals) and by suitable rearrangement of
the polynomial coefficients, one function may be transformed into another,
and some
ambiguity in the presentation of the final answer remains\footnote{Even when
only scalar integrals are present, the form of the polynomial coefficient
depends on how the Gram determinants have been cancelled.}.
To generate and simplify our results,
we have made repeated use of the algebraic manipulation packages
FORM and MAPLE.

We choose to work with all particles in the physical channel corresponding
to $e^+e^-$ annihilation where all quarks are in the final state,
$\gamma^* \to q\bar q Q\bar Q $.
The momenta are labeled as,
\begin{equation}
\gamma^*(p_{1234}) \to q(p_1) + \bar q(p_2) + Q(p_3) + \bar Q(p_4).
\end{equation}
Using momentum conservation, we can systematically eliminate the
photon momentum in favour of the four massless quark momenta.

For quarks $q$ and $Q$,
the electric charges are denoted $e_q$ and $e_Q$ while the colours of
the quarks are labeled by $c_i$, $i=1,\ldots,4$.
The colour structure of the matrix element at tree-level $(n=0)$ and
one-loop $(n=1)$ is rather simple and we have,
\begin{eqnarray}
\lefteqn{\M^{(n)} =  \epsilon^\mu \M_\mu^{(n)} = \frac{eg_s^2}{2}
\left(\frac{g_s}{4\pi}\right)^{2n}}\nonumber
\\
&\times & \left\{
\delta_{c_1c_4}\delta_{c_3c_2}
\left(\A^{(n)}_1(1,2)+\A^{(n)}_1(3,4)+\frac{\delta_{qQ}}{N}
\left(\A^{(n)}_2(1,4)+\A^{(n)}_2(3,2)\right)
\right) \right .
\nonumber \\
&-&\left . \delta_{c_1c_2}\delta_{c_3c_4}\left(
\frac{1}{N}
\left(\A^{(n)}_2(1,2)+\A^{(n)}_2(3,4)
\right)
+\delta_{qQ}
\left(\A^{(n)}_1(1,4)+\A^{(n)}_1(3,2)\right) \right)
\right\},
\end{eqnarray}
where $N$ is the number of colours and $\delta_{qQ} = 1$ for identical
quarks and zero otherwise.
The arguments of $\A_i$ indicate which quark line is attached to the
vector boson and hence which quark charge that function is proportional
to.
At lowest order,
\begin{equation}
\A^{(0)}_1(i,j)=\A^{(0)}_2(i,j),
\end{equation}
while at one-loop we find,
\begin{eqnarray}
\A^{(1)}_1(i,j) &=& N A^{(1)}_C(i,j) -
\frac{1}{N} \left(2A^{(1)}_A(i,j)+A^{(1)}_B(i,j)\right), \\
\A^{(1)}_2(i,j) &=& N \left(A^{(1)}_C(i,j)-A^{(1)}_A(i,j)\right)
-\frac{1}{N} \left(A^{(1)}_A(i,j)+A^{(1)}_B(i,j)\right).
\end{eqnarray}
The functions $A^{(1)}_\alpha(i,j)$, $\alpha = A,B,C$ represent the
contributions of the three gauge invariant sets of Feynman diagrams
shown in Fig.~1 where the photon couples to the quark-antiquark
pair $i,j$.
The set of diagrams containing closed fermion triangles
do not contribute by Furry's theorem.
Note that we choose to include the contribution from the fermion loop
in the fourth diagram of Fig.~1(c) which is proportional to the number of
flavours $N_F$ in the leading colour part $\A_C$.

\begin{figure}\vspace{18cm}
\includegraphics{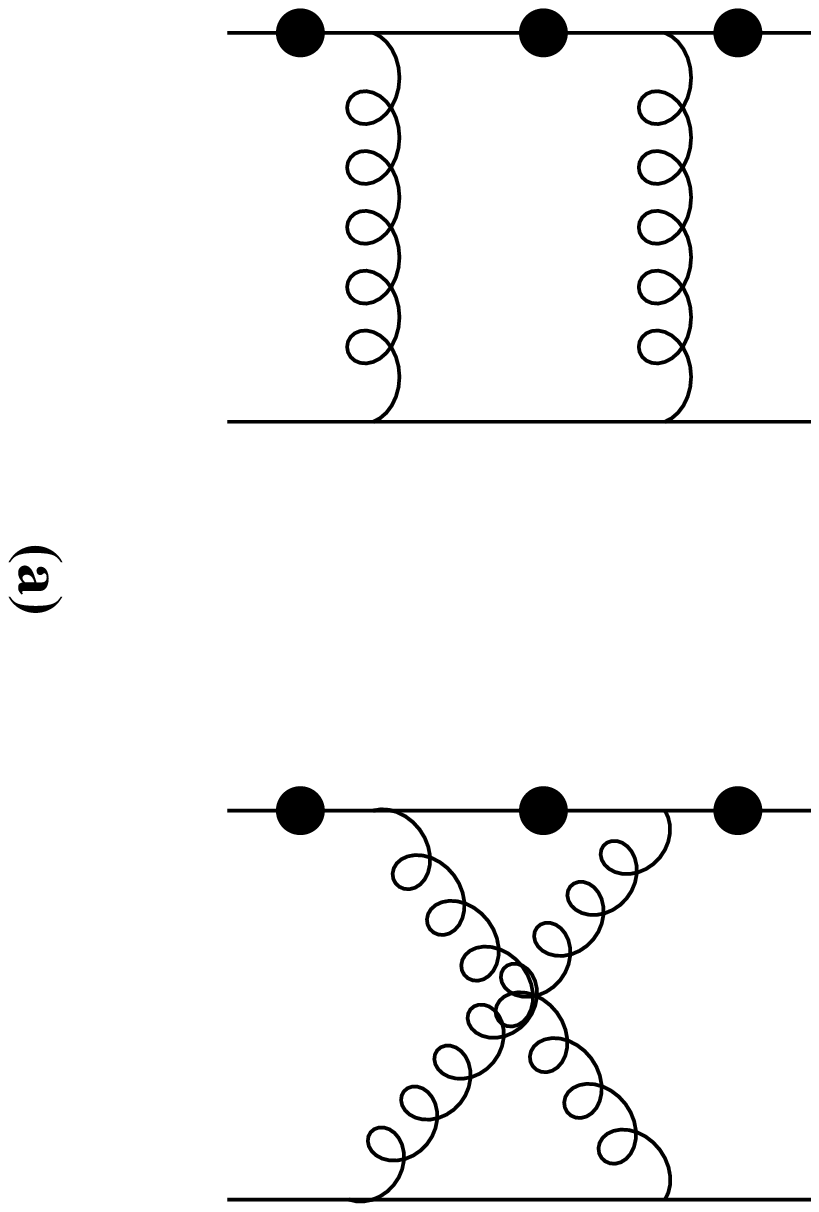}
\includegraphics{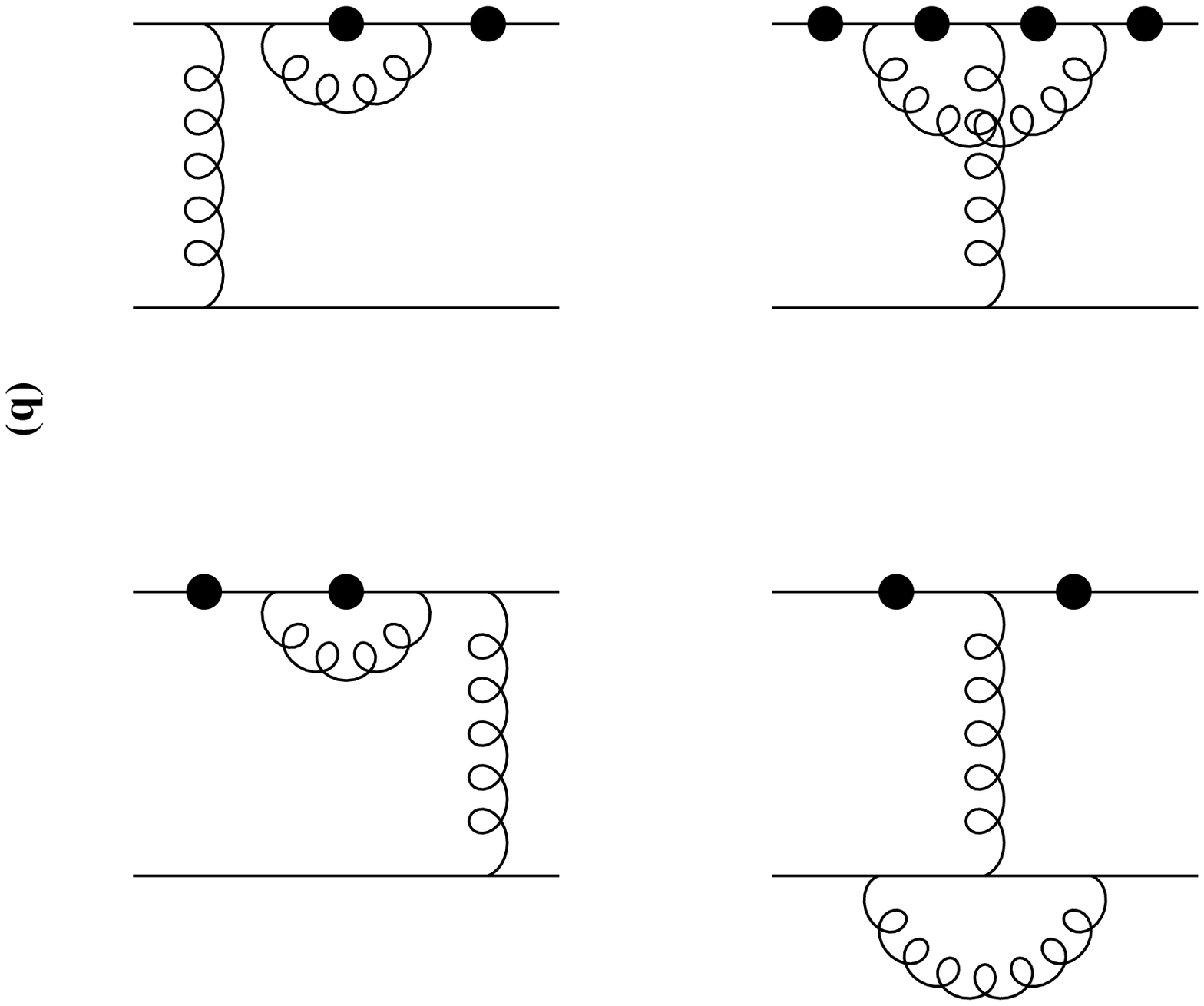}
\includegraphics{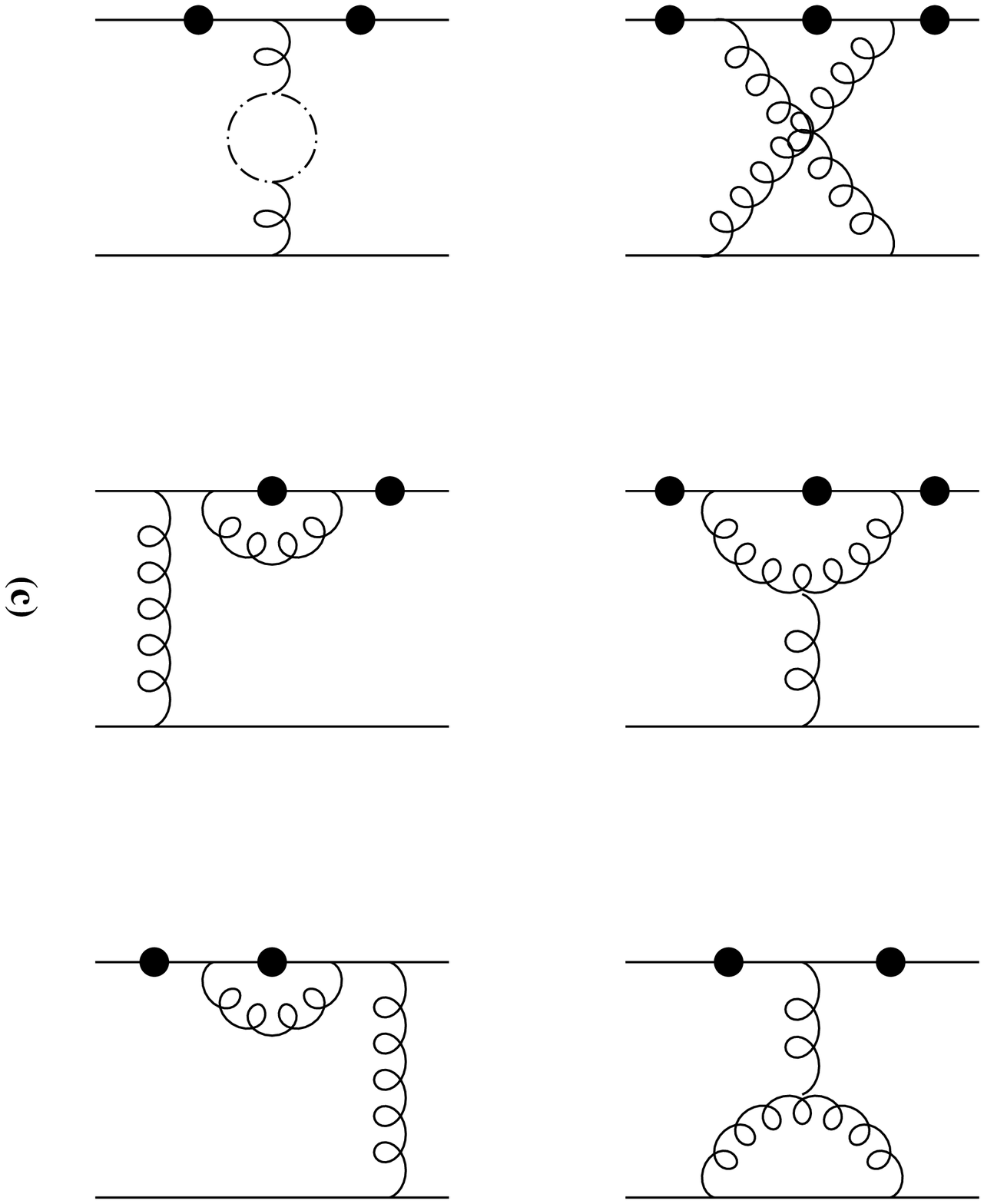}
\caption[]{The classes of Feynman diagrams relevant for the
different colour structures.
The solid circle indicates the possible positions for attaching the off-shell
photon to the quark-antiquark
pair $i,j$.
Group (a) contributes to $A^{(1)}_A(i,j)$, (b) to $A^{(1)}_B(i,j)$ and
group (c) to the leading colour amplitude $A^{(1)}_C(i,j)$.
Diagrams with self-energy corrections on the external lines are zero in
dimensional regularisation
and have been omitted. }
\end{figure}

The squared matrix elements are straightforwardly obtained
at leading order \cite{ERT}.
Because the matrix element contracted with the photon momentum
vanishes by current conservation, the
sum over the polarisations of the gauge boson may be performed
in the Feynman gauge,
\begin{equation}
\sum_{\rm spins} \epsilon^\mu \epsilon^{*\nu} = -g^{\mu\nu}.
\end{equation}
Hence,
\begin{eqnarray}
\lefteqn{
\sum_{\rm spins} |\M^{(0)}|^2  = \frac{e^2 g_s^4}{4}(N^2-1)} \nonumber
\\
&\times&\Biggl \{
\biggl(\T(1,2;1,2)+\T(1,2;3,4)\biggr)
+\frac{\delta_{qQ}}{N} \biggl(\T(1,2;1,4) + \T(1,2;3,2) \biggr)\Biggr \}
\nonumber \\
 &&+ (1 \leftrightarrow 3,2 \leftrightarrow 4)
 + \delta_{qQ} (2 \leftrightarrow 4)
 + \delta_{qQ} (1 \leftrightarrow 3),
\end{eqnarray}
where,
\begin{equation}
\T(i,j;k,l) = \sum_{\rm spins}
| A^{(0)\dagger}(i,j) \A^{(0)}(k,l) |.
\end{equation}

The relevant `squared' matrix elements are the interference between
the tree-level and one-loop amplitudes,
\begin{eqnarray}
\lefteqn{
\sum_{\rm spins} 2 |\M^{(0)\dagger}\M^{(1)}|  = \frac{e^2g_s^4}{4}
\left(\frac{\alpha_sN}{2\pi}\right)(N^2-1)}\nonumber
\\
&\times&\Biggl \{
\biggl[ \biggl(\L_C(1,2;1,2)+\L_C(1,2;3,4)\biggr)
\nonumber \\
&&~~~~~~~-\frac{1}{N^2} \biggl( 2\L_A(1,2;1,2)+2\L_A(1,2;3,4)
+\L_B(1,2;1,2)+\L_B(1,2;3,4)\biggr)\biggr]
\nonumber \\
&&+ \delta_{qQ}
\biggl[ \frac{1}{N}\biggl(\L_C(1,2;1,4)+\L_C(1,2;3,2)
-\L_A(1,2;1,4)-\L_A(1,2;3,2)\biggr)
\nonumber \\
&&~~~~~~~-\frac{1}{N^3} \biggl( \L_A(1,2;1,4)+\L_A(1,2;3,2)
+\L_B(1,2;1,4)+\L_B(1,2;3,2)\biggr) \biggr] \Biggr \}
\nonumber \\
 &&+ (1 \leftrightarrow 3,2 \leftrightarrow 4)
 + \delta_{qQ} (2 \leftrightarrow 4)
 + \delta_{qQ} (1 \leftrightarrow 3),
\end{eqnarray}
with,
\begin{equation}
\L_\alpha(i,j;k,l) = \sum_{\rm spins}
| A^{(1)\dagger}_\alpha(i,j) \A^{(0)}(k,l) |.
\end{equation}
Using the symmetry properties of the Feynman diagrams, we find that,
\begin{equation}
\L_\alpha(1,2;3,2) =  \L_\alpha(1,2;1,4)(p_1 \leftrightarrow p_2,
p_3 \leftrightarrow p_4),
\end{equation}
so that the `squared' matrix elements are described by 9 independent
$\L_\alpha$.

In computing the `square', the loop momentum $\ell^\mu$
always appears as
either $\ell^2$ or $\ell.p$.
We therefore systematically reduce the
number of loop momenta appearing in the numerator by simple rewriting,
$$
\frac{\ell.p}{\ell^2 (\ell+p)^2} =
\frac{(\ell+p)^2-\ell^2-p^2}{2\ell^2 (\ell+p)^2}.
$$
The only time that we are unable to do this occurs when the loop
momentum is contracted with the `wrong' momentum, i.e., one that
does not appear in any of the propagators.
For the pentagon diagrams, all momenta are involved in the propagators
and
we can always rewrite the integral as a combination of box
tensor-integrals.
Dividing through in this way, we find box (triangle) tensor-integrals
with at most three (two) loop momenta in the numerator.

The simplified tensor integrals can now be expressed in terms of scalar
integrals
using a momentum decomposition.
We choose the natural momentum as in \cite{PV} rather than
the reciprocal momentum basis of \cite{OV,EGY,Signer}.
The coefficients of the scalar integrals are obtained by solving systems
of linear equations and the usual problem is the occurrence of Gram
determinants.
Typically, for a rank $n$ tensor $m$-point diagram, we can obtain
an $(m-1) \times (m-1)$ Gram determinant raised to the $n$th power
in the denominator.
Since the physical process does not usually possess singularities
near the edge of phase space where the Gram determinants vanish,
one obtains large cancellations between terms and potential numerical
inaccuracies.
For `squared' one-loop matrix elements the number of terms is expected
to be large, thereby exacerbating the numerical problem.
However, since these Gram determinant singularities are unphysical,
it should prove possible to arrange the scalar integrals in
combinations that are finite in the limit that the Gram determinant
vanishes.
The matrix elements, written in terms of these finite combinations,
are then obviously finite  and numerically stable
in the same limits.
The explicit results of
Bern, Kosower and Dixon \cite{BDK5g,BDK2q3g} and
Kunszt, Signer and Trocsanyi \cite{KST4qg,Signer}
for the five-parton one-loop amplitudes already show the introduction of such
functions.
For example, Bern, Kosower and Dixon  introduce the functions
$$
L_0(r) = \frac{\log(r)}{1-r},~~~~~~~~L_1(r) = \frac{L_0(r)+1}{1-r},
$$
which are finite as $r\to 1$.
Furthermore, the helicity coefficient of these functions is usually
rather simple so one might expect that the `squared' matrix element
coefficient is not enormous.
We have therefore assembled a collection of finite functions
relevant to the process at hand \cite{CGM}
and organised the $\L_\alpha$ as
a linear combination of these functions where the coefficients are
polynomials
of the generalised Mandelstam invariants,
$$
s_{ij}= (p_i+p_j)^2,~~~~~s_{ijk} =(p_i+p_j+p_k)^2,~~~~~
s_{ijkl} = (p_i+p_j+p_k+p_l)^2.
$$
As mentioned earlier, these fuctions are natural in the sense
that they are obtained by
differentiating the scalar integral with respect to the external
kinematic parameters or by evaluating the integral in higher dimensions.
However, these finite combinations are not linearly
independent (unlike the raw scalar integrals), and by explicitly cancelling
part of the polynomial coefficient against the Gram determinant
it is possible to shift terms from one function to another.
Therefore an
ambiguity in the presentation of the final answer remains.

As mentioned earlier, we use dimensional regularisation and work in
$D=4-2\e$ dimensions.
It is straightforward to remove the infrared and ultraviolet singularities
poles from the $\L_\alpha$ since they are proportional to the tree-level
amplitudes,
\begin{eqnarray}
\L_A(1,2;i,j) &=&
\left( +\frac{\P_{13}}{\e^2} -\frac{\P_{14}}{\e^2}
-\frac{\P_{23}}{\e^2}+\frac{\P_{24}}{\e^2}\right)
\T(1,2;i,j) + \Lhat_A(1,2;i,j),  \\
\L_B(1,2;i,j) &=&
\left(-\frac{\P_{12}}{\e^2}-\frac{\P_{34}}{\e^2} -\frac{3\P_{34}}{\e}\right)
\T(1,2;i,j) + \Lhat_B(1,2;i,j),  \\
\L_C(1,2;i,j) &=&  \left(
-\frac{\P_{14}}{\e^2}-\frac{\P_{23}}{\e^2}
+\frac{2}{3}\frac{\P_{34}}{\e}-\frac{2N_F}{3N}\frac{\P_{34}}{\e}\right)
\T(1,2;i,j) + \Lhat_C(1,2;i,j),
\end{eqnarray}
where we have introduced the notation,
\begin{equation}
\P_{ij} = \left(\frac{4\pi\mu^2}{-s_{ij}}\right)^{\e}
\frac{\Gamma^2(1-\e)\Gamma(1+\e)}{\Gamma(1-2\e)}.
\end{equation}

In physical cross sections, this pole structure must cancel with
the infrared poles from the $\gamma^* \to q\bar qQ\bar Q +g$
process and those generated by ultraviolet renormalisation.
This pole structure is in agreement with the expectations
of ref.~\cite{GG} and reproduces that given in \cite{Sig4qp,Signer}.

The finite functions $\Lhat$ can be written in the following
symbolic way,
\begin{equation}
\Lhat  =
\sum_i P_i(s) {\rm L}_i,
\end{equation}
where $P_i(s)$ is a polynomial of the invariant masses $s_{ij}$
and the ${\rm L}_i$ are linear combinations of scalar loop integrals.
Rather than have the Gram determinants present in the polynomial coefficients
as in the conventional method \cite{PV,OV,EGY,Signer}),
they are absorbed into the ${\rm L}_i$ which are constructed
to be well behaved in the limit that the Gram determinant vanishes.
The advantage is better numerical stability while the disadvantage is that
there are rather more functions (than the raw scalar integrals in
$4-2\e$ dimensions).  Furthermore, since the functions are not linearly
independent the polynomial coefficients are not unique.
Typically, the $P_i(s)$ contains ${\cal O}(30)$ terms (roughly
the same size as the tree level result) while the summation runs
over ${\cal O}(30)$ functions.
Although the expreassions for the
individual $\Lhat_\alpha$ are in closed form, they are
still rather lengthy and each
contains several hundred terms.
We have therefore  constructed a FORTRAN subroutine
detailing the $P_i(s)$ and
evaluating the finite one-loop contribution for a given phase space point.
This will act as an ingredient in calculating
the full next-to-leading order $e^+e^- \to 4$~jet rate once the
necessary sub-leading colour
contributions to the $\gamma^* \to q\bar q gg$ matrix elements are known
via numerical combination with the tree level
$\gamma^* \to 5$~parton
processes \cite{HZ,BGK,FGK}.

To illustrate the finite groupings ${\rm L}_i$ and the potential
numerical instabilities, we show the
functions associated with the triangle graph with
three massive external legs.
The scalar triangle integral with momenta $p_{12}$ and $p_{34}$
flowing out (and $p_{1234}$ flowing in) can be written \cite{3pt,EGY},
\begin{equation}
\Lc{0}(p_{12},p_{34})
=\frac{1}{\sqrt{-\Delta_3}}
\left(\log(a^+a^-)\log\left(\frac{1-a^+}{1-a^-}\right)
+2\Li2(a^+)-2\Li2(a^-)\right),
\end{equation}
where $\Li2$ is the usual dilogarithm function and
$a^\pm $ are two roots of a quadratic equation,
\begin{equation}
a^{\pm}=\frac{s_{1234}+s_{34}-s_{12} \pm  \sqrt{-\Delta_3} }{2 s_{1234}},
\end{equation}
and,
\begin{equation}
\Delta_3 =-s_{1234}^2-s_{12}^2-s_{34}^2
+2 s_{1234} s_{12}+2 s_{1234} s_{34}+2 s_{12} s_{34}.
\end{equation}
The functions,
\begin{eqnarray}
\Lc{1S}(p_{12},p_{34})&=&
\frac{1}{2\Delta_3}
\Biggl(2s_{1234}s_{12}s_{34}\Lc{0}(p_{12},p_{34})\\
&&
    -\left
(s_{12}(s_{1234}+s_{34}-s_{12})\log\left(\frac{s_{1234}}{s_{12}}\right)
+s_{34}(s_{1234}+s_{12}-s_{34})\log\left(\frac{s_{1234}}{s_{34}}\right)
\right )\Biggr),\nonumber  \\
\Lc{2S}(p_{12},p_{34})&=&
\frac{1}{4\Delta_3}
\Biggl(2 s_{1234}s_{12}s_{34}\Lc{1S}(p_{12},p_{34})
 -\frac{1}{3}s_{1234}s_{12}s_{34}
\\
&&
    -\frac{1}{6}\left
(s_{12}^2(s_{1234}+s_{34}-s_{12})\log\left(\frac{s_{1234}}{s_{12}}\right)
+s_{34}^2(s_{1234}+s_{12}-s_{34})\log\left(\frac{s_{1234}}{s_{34}}\right)
  \right )\Biggr), \nonumber
\end{eqnarray}
also appear and are related to the scalar triangle integral in $D=6-2\e$ and
$D=8-2\e$ dimensions  \cite{CGM}.
Throughout, the arguments of the
functions  describe the momenta flowing out of all but one of the vertices,
while the final momentum is determined by momentum conservation.
For triangle integrals it is also convenient to introduce the additional
functions corresponding to integrals in $4-2\e$ dimensions with
Feynman parameters in the numerator \cite{CGM}, for example,
\begin{eqnarray}
\Lc{1}(p_{12},p_{34})&=&
\frac{1}{\Delta_3}
\Biggl(s_{12}(s_{1234}-s_{12}+s_{34})\Lc{0}(p_{12},p_{34}) \nonumber \\
&&
    -(s_{1234}-s_{12}-s_{34})\log\left(\frac{s_{1234}}{s_{34}}\right)
            -2s_{12}\log\left(\frac{s_{1234}}{s_{12}}\right)\Biggr ),\\
\Lc{2}(p_{12},p_{34})&=&
\frac{1}{2\Delta_3}       \Biggl(
2s_{34}(s_{1234}+s_{12}-s_{34})\Lc{1}(p_{12},p_{34})
+s_{12}(s_{1234}+s_{34}-s_{12})\Lc{1}(p_{34},p_{12}) \nonumber \\
&&\qquad\qquad
 -s_{12}s_{34} \Lc{0}(p_{12},p_{34})
        - s_{34}\log\left(\frac{s_{1234}}{s_{34}}\right)
+s_{12}+s_{34}-s_{1234}
\Biggr ).
\end{eqnarray}
As $\Delta_3 \to 0$, all of these functions are finite and combine dilogarithms
and logarithms in a highly non-trivial way.
For example,
\begin{eqnarray}
\lim_{\Delta_3 \to 0} \Lc{1S}(p_{12},p_{34}) &=&
\frac{1}{12s_{12}s_{34}s_{1234}}
\Biggl(s_{12}^2(s_{1234}+s_{34}-s_{12})\log\left(\frac{s_{1234}}{s_{12}}\right)
\nonumber \\ &&\qquad
+
 s_{34}^2(s_{1234}+s_{12}-s_{34})\log\left(\frac{s_{1234}}{s_{34}}\right)
+2s_{12}s_{34}s_{1234} \Biggr).
\end{eqnarray}
By keeping terms proportional to $\Delta_3$ (and higher),
the limit can be approached with arbitrary precision.

\begin{figure}\vspace{8cm}
\includegraphics{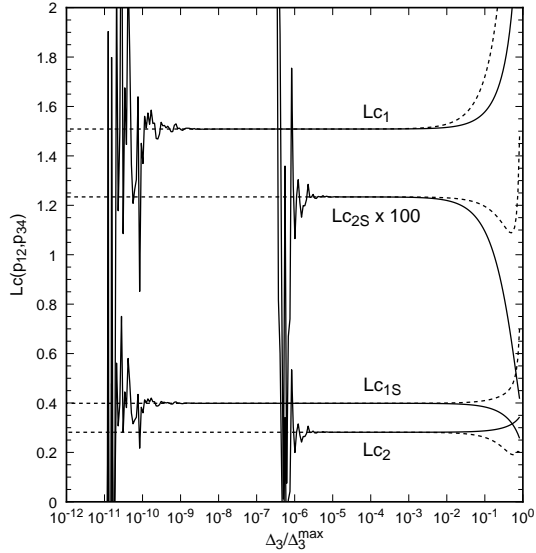}
\caption[]{The finite functions for the triply massive triangle graph
with $s_{1234} = 1$ and $s_{12} = 0.2$ as a function of
$\Delta_3/\Delta_3^{{\rm max}}$ where $\Delta_3^{{\rm max}} =
-(s_{1234}-s_{12})^2$.
The dashed lines show the
approximate form for the function in the limit $\Delta_3 \to 0$,
retaining only the first term of the
Taylor expansion.}
\label{fig:lc}
\end{figure}

Numerically, the instability due to the Gram determinant is easy to see
if we restrict ourselves to a specific phase space point,
$s_{1234} = 1,~~s_{12} = 0.2$ and letting $s_{34}$ vary in such
a way that $\Delta_3 \to 0$.
This corresponds to $s_{34} \to 0.135$ which is neither the soft nor collinear
limit
and there should be no kinematic singularity.
The behaviour of the various triangle functions is illustrated in
Fig.~\ref{fig:lc}
as a function of $\Delta_3/\Delta_3^{{\rm max}}$ where
$\Delta_3^{{\rm max}} =
-(s_{1234}-s_{12})^2$
For a given numerical precision, $acc$,
the numerical problems typically occur
when $\Delta_3/\Delta_3^{{\rm max}} \sim (acc)^{1/N}$ where $N$ is the
number of Gram determinants in the denominator of the function.
Since the scalar integral already contains $\Delta_3^{-1/2}$, for an
intrinsic numerical precision of roughly $10^{-14}$, the $\Lc{1}$ and $\Lc{1S}$
functions break down at  $\Delta_3/\Delta_3^{{\rm max}} \sim {\rm few~}
\times 10^{-10}$
while the $\Lc{2}$ and $\Lc{2S}$ functions with effectively $2.5$ inverse
powers of $\Delta_3$ break down at $\Delta_3/\Delta_3^{{\rm max}}
\sim {\rm few~} \times 10^{-6}$.
Other phase space points yield a similar behaviour.
With adaptive Monte Carlo methods such as VEGAS \cite{VEGAS}, once the
singular region is found, more and more Monte Carlo points are generated
there and the result will be unreliable.
However, for $\Delta_3 < 10^{-5}
\Delta_3^{{\rm max}}$ we see that the approximate form obtained by making a
Taylor expansion
about $\Delta_3 = 0$ and keeping only the constant term  works well.

Similar groupings for the box and pentagon graphs (with their corresponding
Gram determinants) can be obtained by considering the scalar integrals in
higher dimensions or with Feynman parameters in the numerator \cite{CGM}.
This approach is easily generalisable to other processes with more
general kinematics.

To summarize,
we have performed the first calculation of the one-loop `squared'
matrix elements for the $\gamma^* \to q\bar q Q\bar Q $ process
keeping all orders in the number of colours.
Throughout conventional dimensional regularisation has been employed
as well as grouping the Feynman diagrams according to the
colour structure.
However, in order to deal with the Gram determinants, we have
expressed the formfactors in terms of functions (${\rm L}_i$)
which group together dilogarithms, logarithms and constants from the
basic scalar integrals in a way that
is well behaved as the Gram determinants vanish.
These groupings are derived by differentiating the scalar integrals
with respect to the external kinematic factors or by considering
them in higher dimensions.
The resulting expressions are still rather long,
due partly to the number of functions and partly to the
size of the tree-level matrix elements,
but numerically stable and ready to be implemented in
next-to-leading order
Monte Carlo calculations of the
$e^+e^- \to 4~{\rm jet}$,
$e^\pm p \to e^\pm + 3~{\rm jet}$ and
$p\bar p \to V + 2~{\rm jet}$  processes.

{\noindent {\bf Acknowledgements}}

We thank Walter Giele, Eran Yehudai, Bas Tausk and Keith Ellis
for collaboration in the earlier stages of this work
and John Campbell for numerous comments and suggestions in the latter stages.
DJM thanks the UK Particle Physics and Astronomy Research Council
for the award of a research studentship.

\end{document}